\documentclass[]{aa}
\usepackage{graphics}
\usepackage{graphicx}
\usepackage{txfonts}
\usepackage{natbib}
\bibpunct{(}{)}{;}{a}{}{,}

\def\apj{ApJ}                 
\def\apjl{ApJ Lett.}                
\def\apjs{ApJS}

\def\aap{A\&A}

\def\mnras{MNRAS}

\def\prl{Phys.~Rev.~Lett.}

\def\solphys{Sol.~Phys.}

\def\zap{ZAp}

\def\nphysa{Nucl.~Phys.~A}

\def\lsol{L$_{\odot}$}
\def\rsol{R$_{\odot}$}
\begin{document}      
%
   \title{Low abundances of heavy elements in the solar outer layers: comparisons of solar models with helioseismic inversions} 
%
 
        \titlerunning{Low abundances of heavy elements in the solar outer layers}  
 
   \author{
M. Castro \ 
\inst{1}, 
S. Vauclair 
\inst{1},
\and O. Richard
\inst{2}
}
 
   \offprints{M. Castro}

   \institute{Laboratoire d'Astrophysique de Toulouse et Tarbes - UMR 5572 - Universit\'e Paul Sabatier Toulouse III - CNRS, 14, av. E. Belin, 31400 Toulouse, France
\and GRAAL - UMR 5024 - Universit\'e Montpellier II - CC072, Place E. Bataillon, 34095 Montpellier, France} 

\mail{mcastro@ast.obs-mip.fr}
   
\date{Received \rule{2.0cm}{0.01cm} ; accepted \rule{2.0cm}{0.01cm} }

\authorrunning{Castro et al.}

\abstract
{Recent solar photospheric abundance analyses have led to a significant reduction of the metal abundances compared to the previous determinations. $(Z/X)_S$ is 30\% smaller, down to 0.0165. The solar models computed with standard opacities and diffusion processes using these new abundances give poor agreement with helioseismic inversions for the sound-speed profile, the surface helium abundance, and the convective zone depth.}
{We attempt to obtain a good agreement between helioseismic inversions and solar models that present the ``old mixture'' in the interior and new chemical composition in the convective zone. To reach this result, we assume an undermetallic accretion at the beginning of the main sequence.}
{We compute solar models with the Toulouse-Geneva Evolution Code, in which we simulate an undermetallic accretion in the early stages of the main sequence, in order to obtain the new mixture in the outer convective zone. We compare the sound-speed profile, the convective zone depth, and the surface helium abundance with those deduced from helioseismology.}
{The model with accretion but without any mixing process inside is in better agreement with helioseismology than the solar model with the new abundances throughout. There is, however, a spike under the convective zone which reaches 3.4\%. Furthermore, the convective zone depth and the surface helium abundance are too low. Introducing overshooting below the convective zone allows us to recover the good convective zone radius and the addition of rotation-induced mixing and tachocline allows us to reconcile the surface helium abundance. But in any case the agreement of the sound-speed profile with helioseismic inference is worse than obtained with the old abundances.}
{}  


\maketitle
                                                                                                                                               
\section{Introduction} 

The standard solar models that include the metal abundances as given by \citet{GN93} (hereafter GN93) and the current physics updates \citep[new opacity tables and equations of state from OPAL, and element diffusion and settling, e.g.,][]{richard04} are in very good agreement with helioseismology. The square of the sound speed $c^2$ derived from this standard solar model agrees with the values obtained from helioseismology within 0.5\% \citep{JCD96}. The recent solar photospheric abundances derived by \citet{asplund05} (hereafter Asp05) revise the CNO and Ne abundances downwards by more than 30\% compared to the earlier determinations. The solar photospheric $Z/X$ is reduced to $0.0165 \pm 10\%$ and $Z$ to $0.0122$, whereas the ``old'' mixture gave a higher $Z/X$ of 0.0245. New solar standard models including the Asp05 mixture lead to a large disagreement with the helioseismically inferred sound-speed profile, convective zone depth, and surface helium abundance $Y_S$.

Several research groups have explored different ways to try restoring agreement between solar models and helioseismology. \citet{bahcall05a} showed that an increase of 21\% of the OPAL opacities could reconcile the new solar abundances with the helioseismic measurements of the depth of the convective zone. The uncertainties in the OPAL96 opacity tables however are estimated to be no more than 4\%. The uncertainties on the element diffusion velocities below the convective zone are larger and can reach 35\% for Fe and O, the two main contributors to the opacity in this region \citep{montalban06}. An increase of the diffusion coefficient by a factor of 1.5 to 2 combined with a small increase of the opacity at the bottom of the convective zone provides better agreement with helioseismology for the depth of the convective zone and the $c^2$ profile \citep{montalban04,basuantia04}, but in this case $Y_S$ has to be decreased below the value determined by the inversion of solar oscillations. The attempt by \citet{guzik05} to increase the diffusion coefficients for helium and the other elements separately to fit the Asp05 mixture in the outer layers and recover the GN93 mixture at the bottom of the convective zone leads to the same problem. \citet{turck04} showed that the new oxygen measurement of the Sun from \citet{asplund05} was in accordance with the observations of extragalactic HII regions, Magellanic clouds, and other clusters. They updated their predicted neutrino fluxes using the new composition and they obtained a good agreement with the detected values. \citet{antiabasu05} and \citet{bahcall05b} computed models with an arbitrarily increased Ne/O ratio and showed that they could be in agreement with helioseismology. At the same time, \citet{draketesta05} assessed an higher Ne/O ratio for the Sun than the one adopted in \citet{asplund05} and consistent with the results of \citet{antiabasu05} and \citet{bahcall05b}, by observing several cool nearby stars in X-ray spectra. However, new measurements by analyzing the solar active region spectra \citep{schmelz05} and the extreme ultraviolet emission lines of Ne\small{IV-VI} \normalsize and O\small{III-V} \normalsize ions in the solar corona \citep{young05}, are in agreement with the Ne/O ratio value of the new abundances.

Here we compute solar models with the assumption that undermetallic accretion occurred in the very early stages of the solar life on the main sequence. A similar idea was also proposed in \citet{guzik05} without computing any model. We suppose that, during the process of planetary formation, the young Sun was surrounded by metal-poor matter while metals were concentrated in planetesimals, and that an important part of this metal-poor matter fell onto it. We do not justify this idea here by any computations of accretion processes: our aim was only to compute ``exotic'' solar models that could be able to reconcile the new solar abundances with helioseismology. We simulated this undermetallic accretion to obtain the Asp05 mixture in the outer convective zone with an increased abundance inside. In this way, we can obtain a metallic difference between the outer layer and the inside larger than possible from diffusion processes only. In these models, we attempted to recover the sound-speed profile, the depth of the convective zone, and the surface helium abundance as given by helioseismology. 

In Sect. 2, we discuss the new Asp05 abundances and their influence on the sound-speed profile. In Sect. 3, the models with accretion are presented. The effects of overshooting below the convective zone and mixing induced by the tachocline and by rotation are respectively discussed in Sects. 4 and 5. The conclusions are given in Sect. 6.

\section{Solar models with the new Asp05 mixture}

The models are computed with the Toulouse-Geneva Stellar Evolution Code (TGEC) \citep{richard96}. We use the equation of state tables of OPAL2001 \citep[updated version of][]{rogers96}, the opacities tables of OPAL \citep{rogersiglesias95} completed with the \citet{alex94} low temperature opacities, and the NACRE compilation of nuclear reaction rates \citep{angulo99} with the Bahcall screening routine. The convection treatment is based on the mixing-length theory \citep{bohm58} and the diffusion coefficients are computed as in \citet{paquette86}. \\
 
Model S1 is a standard model, computed with the ``old'' GN93 chemical composition. No mixing other than convection is introduced. The calibration procedure includes two free parameters: the initial mass fraction of helium $Y_0$ and the mixing-length parameter $\alpha$. These parameters are adjusted to obtain the solar luminosity and radius at the solar age: \lsol $= 3.8515 \pm 0.0055 \times 10^{33}$ erg.s$^{-1}$ \citep{guenther92}, \rsol $= 6.9575 \pm 0.0024 \times 10^{10}$ cm \citep[mean observed value :][]{allen76,brown98}, and $t_{\odot} = 4.6$ Gyrs. The characteristics of model S1 are presented in Table \ref{characmodels}. 

\begin{figure}
\begin{center}
\includegraphics[angle=0,totalheight=\columnwidth,width=\columnwidth]{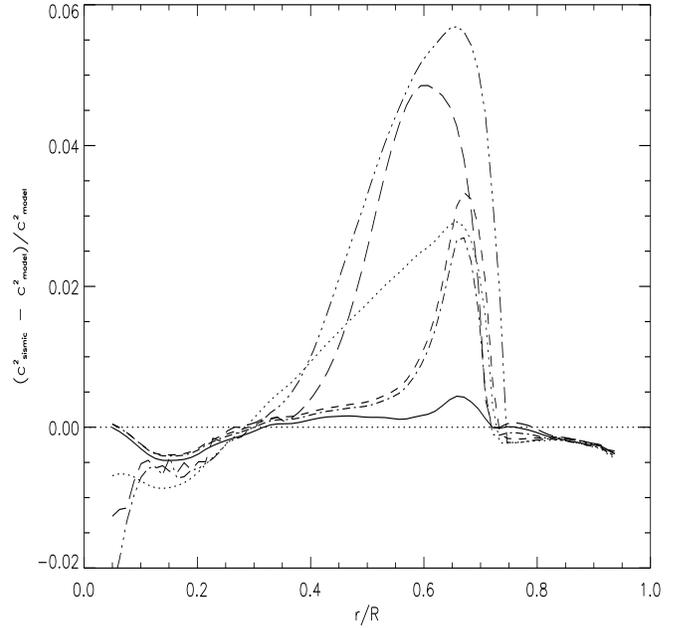}
\end{center}
\caption{Comparison between the profile of sound velocity of the different models (S1 : solid line, S2 : dotted line, S3 : dashed line, S4 : dot-dashed line, S5 : three dot-dashed line, S6 : long dashed line) and the one deduced from the helioseismology} 
\label{ComparBasu}
\end{figure}

This model is in good agreement with helioseismology. The comparison of the sound velocity computed in this model with the one deduced from helioseismology by \citet{basu97} is presented as the solid line in Fig. \ref{ComparBasu}. The relative difference between model S1 and the seismic inversions is always smaller than 0.5\%. The depth of the convective zone is also consistent with helioseismology: we find $r_{cz}/R_{\odot} = 0.712$ compared to $r_{cz}/R_{\odot} = 0.713 \pm 0.001$ \citep{basuantia97}. The helium abundance is 0.240, slightly smaller than the value obtained from seismic inversions, which is typically $0.249 \pm 0.002$ \citep{basuantia97}. This is due to the fact that model S1 does not include any mixing process which could slightly slow down helium settling. \\

\begin{table*}
\caption{Characteristics of the solar models computed with the TGEC.}
\label{characmodels}
\begin{center}
\begin{tabular}[h!]{c|c|c|c|c|c|c|c}
\hline
 & \multicolumn{2}{|c|}{Calibration} & \multicolumn{4}{|c|}{External} & Convective \\
 & \multicolumn{2}{|c|}{parameters} & \multicolumn{4}{|c|}{parameters} & zone \\
\hline
 & $\alpha$ & $Y_0$ & $L$ ($10^{33}$ erg.s$^{-1}$) & $R$ ($10^{10}$cm) & $Y_S$ & $Z/X$ & $r_{cz}/R_{\odot}$ \\
\hline
Model S1 & 1.8104 & 0.2716 & 3.8517 & 6.9575 & 0.240 & 0.0244 & 0.713 \\
(GN93) & & & & & & & \\
\hline
Model S2 &  1.6556 &  0.2562 &  3.8511 &  6.9576 &  0.223 &  0.0164 &  0.730 \\
(Asp05) & & & & & & & \\
\hline 
Model S3 & 1.6338 & 0.2720 & 3.8514 & 6.9586 & 0.240 & 0.0165 & 0.732 \\
(GN93 & & & & & & & \\
+ accr) & & & & & & & \\
\hline
Model S4 & 1.6344 & 0.2721 & 3.8519 & 6.9552 & 0.243 & 0.0166 & 0.712 \\
(GN93 & & & & & & & \\
 +accr +ov) & & & & & & & \\
\hline
Model S5 & 1.5101 & 0.2691 & 3.8514 & 6.9592 & 0.249 & 0.0167 & 0.751 \\
(GN93 & & & & & & & \\
 +accr +mix) & & & & & & & \\
\hline
Model S6 & 1.5495 & 0.2705 & 3.8515 & 6.9593 & 0.249 & 0.0165 & 0.712 \\
(GN93 & & & & & & & \\
 +accr +mix +ov) & & & & & & & \\
\hline
\end{tabular}
\end{center}
\end{table*}  

Model S2 has been computed using the Asp05 mixture throughout. Its characteristics are presented in Table \ref{characmodels} and the sound velocity in Fig. \ref{ComparBasu}. The discrepancy between the model and the seismically derived sound velocity is much larger than for model S1: it goes up to 3\%. The initial Z/X is lowered by 33\%, compared to model S1: the new initial value is 0.0183 instead of 0.0272. More precisely, the CNO and Ne mass fractions have decreased by 36\%. The quantitative analysis of \citet{turcotte98} for the GN93 mixture shows that the main contributors to the radiative opacity below the solar convective zone are O, Fe, H, Ne, and, with smaller contributions, N and C. Hence, the reduction of the metal mass fraction and the decrease of O and Ne abundances induce important changes in the opacity related to the Asp05 mixture: the opacity at the bottom of the convective zone decreases by 25\% compared to the GN93 case. As the radiative temperature gradient $\nabla_{rad}$ is proportional to the opacity $\kappa$, the new convective zone is shallower ($r_{cz}/R_{\odot} = 0.729$). Furthermore, the helium abundance in the convective zone is lower than the helioseismic value ($Y_S = 0.223$).

\section{A new undermetallic model}

The problem of new solar models seems very difficult to solve, as the new determinations of solar abundances, using new techniques of 3D-modelization, are incompatible with the seismic studies of the Sun. On the other hand, the old GN93 solar mixture gave a very good consistency with the helioseismic analysis, within 0.5\%. We decided here to compute new models using the GN93 mixture in the solar interior, to try to remain consistent with helioseismology, but with the Asp05 chemical composition inside the convective zone. This situation cannot be justified in terms of element diffusion only, as the precision on the diffusion velocities does not allow such a large settling \citep{montalban04,montalban06}. For this reason, we assumed accretion of metal-poor gas at the beginning of the main sequence. 

We simulated this accretion in a simplified way, by instantaneously decreasing the metal abundances in the convective zone and increasing hydrogen and helium accordingly, in one of our first main sequence models (aged 74 Myrs). If $\tau_{Z}$ is the factor (smaller than one) by which the abundances of metals are multiplied, the hydrogen and helium abundances are multiplied by $\tau_{XY}$, so that :
\begin{equation}
\tau_{XY} = \frac{1 - (1 - X - Y)*\tau_{Z}}{X + Y}
\end{equation}
In this simplified way, all metals are decreased with the same factor, so that the mixture remains unchanged, while the overall ratio $(Z/X)_S$ reaches the \citet{asplund05} value.

\begin{figure}
\begin{center}
\includegraphics[angle=0,totalheight=\columnwidth,width=\columnwidth]{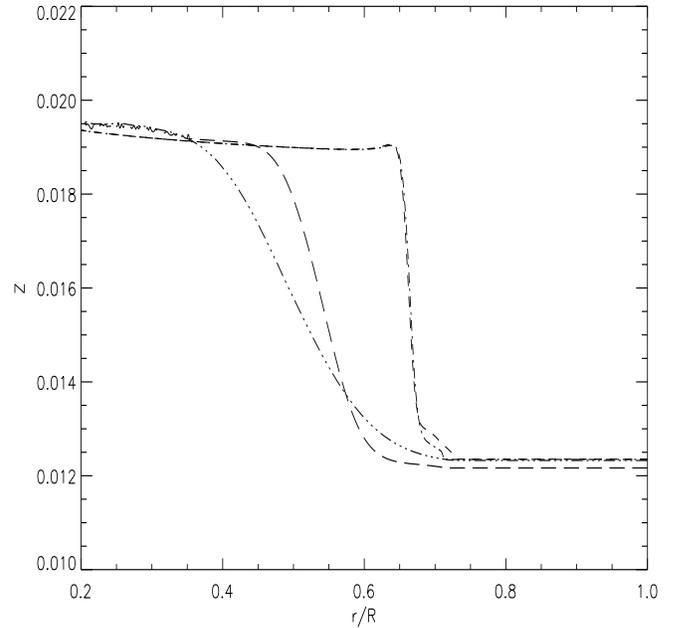}
\end{center}
\caption{Metal profile below the convective zone of models S3 (dashed line), S4 (dot-dashed line), S5 (three dot-dashed line), and S6 (long dashed line)} 
\label{gradmetal}
\end{figure}

Model S3, whose characteristics are also presented in Table \ref{characmodels}, has been computed with $\tau_{Z}=0.5$. The comparison of the sound velocities with the seismic inversions is presented in Fig. \ref{ComparBasu}. The discrepancy is smaller than for model S2, but a spike remains below the convective zone, which reaches 3.4\%. This spike can be due to the fact that the convective zone is too shallow (see Table \ref{characmodels}) and also to the metal gradient induced by accretion and diffusion below the convective zone (see Fig. \ref{gradmetal}). Meanwhile, the helium abundance in the convective zone is still too low. Two kinds of improvements are possible at this stage: introducing an overshooting zone below the convective zone so that the combined mixing reaches the depth deduced by helioseismology ($r_{cz}/R_{\odot}=0.713$), and adding a rotation-induced mixing below the convective zone to smooth down the metal gradient. In the following, we present three new models, one with overshooting (S4), one with rotation-induced mixing (S5), and one with both processes (S6). 

\section{Convective overshooting}

The introduction of overshooting below the convective zone allows us to simulate an overall mixing zone down to the depth deduced from helioseismology. This result is obtained with an overshooting parameter of $0.25 \, \mathrm{H_P}$, where $\mathrm{H_P}$ is the pressure scale height. Due to overshooting, a larger value of the accretion parameter is needed to obtain the Asp05 abundances in the convective zone at the solar age: $\tau_{Z}=0.58$. The characteristics of the resulting model S4 are presented in Table \ref{characmodels}. We can see that the bottom of the convective-plus-overshooting zone is the same as in model S1, namely $r/R_{\odot} = 0.712$. The comparison of the sound velocities with helioseismology is presented in Fig. \ref{ComparBasu} and the metal gradient in Fig. \ref{gradmetal}. The amplitude of the spike under the convective zone is smaller than without overshooting, but it still reaches 2.8\%. The surface helium abundance is still too low.

\section{Rotation-induced mixing}

Mixing below the convective zone was not introduced in the previous models. As a consequence, lithium was not depleted as observed in the Sun. Solar spectroscopic observations compared to abundance determinations in meteorites show that the solar lithium has been depleted by a factor of 160 compared to its initial abundance and that beryllium has not been destroyed by more than 0.09 dex \citep{asplund04}: these features have to be taken into account in consistent solar models. Rotation-induced mixing can deplete lithium as observed and it also smooths down the metal gradient induced by accretion. We computed model S5 (see Table \ref{characmodels}) without overshooting, but with the rotation-induced mixing calibrated to have the suitable lithium destruction at the solar age in question. We used the prescription by \citet{theado03}, with a diffusion coefficient $D_{turb}$, such that:
\begin{equation}
\rho \frac{\partial \bar{c}}{\partial t} = \frac{1}{r^2} \frac{\partial}{\partial r} \left( r^2 \rho D_{turb} \frac{\partial \bar{c}}{\partial r} \right),
\end{equation}
where
\begin{equation}
D_{turb} = \left( C_v + \frac{1}{30 C_h} \right) r \vert U_r \vert = \alpha_{turb} r \vert U_r \vert .
\end{equation}
Here $\vert U_r \vert$ represents the vertical component of the meridional circulation velocity, and $C_h$ and $C_v$ are characteristic parameters for the horizontal and vertical diffusion coefficients:
\begin{eqnarray}
D_h & = & C_h r \vert U_r \vert \\
D_v & = & C_v r \vert U_r \vert .
\end{eqnarray}
$C_h$ and $\alpha_{turb}$ are the free parameters used for the calibration that determines the efficiency of the turbulent motions. We use the values given by \citet{theado02}: $C_h=50000$ and $\alpha_{turb}=6$.

We also include in these computations a tachocline calibrated by two coefficients: the diffusion coefficient at the base of the convective zone $D_{bcz}$ and the thickness of the tachocline $\Delta$. The diffusion coefficient inside the tachocline is given by:
\begin{equation}
D_{tacho} = D_{bcz} \exp \left( \ln 2 \frac{r - r_{bcz}}{\Delta} \right),
\end{equation}
where $r_{bcz}$ is the radius at the base of the convective zone.
These parameters are adjusted to reproduce the observed solar lithium depletion at the solar age: $D_{bcz}=8.0 \times 10^5$ cm$^2$.s$^{-1}$ and $\Delta=0.048 \times 10^{10}$ cm. As mixing reduces element diffusion, a stronger accretion is needed. In model S5, the accretion factor was taken as : $\tau_{Z}=0.37$ and it was introduced in seven time steps. The corresponding metal profile in the present Sun is shown in Fig. \ref{gradmetal}. The comparisons of the sound velocity in model S5, compared to the seismic value, is presented in Fig. \ref{ComparBasu}. \\

The agreement is worse than for previous models. The spike under the convective zone reaches 5.7\% and the difference with the seismic value remains large far below the convective zone, down to $r/R_{\odot}=0.3$. This is due to the spread of the metal gradient in the interior and the consecutive changes in the opacity. Here the convective zone is still too shallow ($r_{cz}/R_{\odot}=0.751$), but the helium abundance is correct in the convective zone ($Y_S=0.249$) as well as the lithium value ($Li/Li_0 = 1/167$). Beryllium is depleted by 0.4 dex, which is too much compared to the observed value. \\

Model S6 has been computed with both overshooting and rotation-induced mixing. The accretion and mixing parameters were modified compared to previous models : $\tau_{Z}=0.48$ introduced in three time steps, $C_h=65000$, $\alpha_{turb}=1$, $D_{bcz}=2.5 \times 10^5$ cm$^2$.s$^{-1}$, and $\Delta=0.048 \times 10^{10}$ cm. The sound-speed profile compared with helioseismology is presented in Fig. \ref{ComparBasu} and the metal gradient in Fig. \ref{gradmetal}. The metal gradient is steeper in model S6 than in model S5 because, thanks to the overshooting, the rotation-induced mixing which reproduces the lithium underabundance has to be less efficient. Here beryllium is only depleted by 0.12 dex. The overshooting allows us to reach the right depth of the convective zone, and the mixing gives the right surface helium abundance, but the sound-speed profile is still in large disagreement with helioseismology.

\section{Conclusion}

The incompatibility between the new solar photospheric abundances of \citet{asplund05} and helioseismology have already been studied by many authors \citep{bahcall05a,montalban04,montalban06,basuantia04,guzik05}. No satisfying solution has been found. The best agreement is obtained by increasing the opacity below the convective zone and the diffusion rates with suitable factors \citep{montalban06}. But the required changes are larger than the accepted uncertainties, and the models still have too low surface helium abundances. In the present paper, we have computed solar models with the GN93 mixture in the solar interior and the Asp05 abundances in the external convective zone. These models could be justified by accretion of metal-poor gas in the early stage of the main-sequence.

Models in which such an accretion is introduced present a better agreement with helioseismology than models using the Asp05 mixture throughout, but there is still a large discrepancy below the convective zone in the sound velocity, and the depth of the convective zone as well as the surface helium abundance are incorrect. The introduction of overshooting below the convective zone allows us to reach a mixing zone as deep as that deduced from helioseismology. Adding rotation-induced mixing allows us to obtain the right surface helium abundance, and also the right lithium depletion. While the introduction of overshooting does not change the sound-speed profile in a significant way, the smoothing and spreading of the metal gradient due to rotation-induced mixing increases the difference with the seismic sound velocity profile in the interior. We tried our best to reconcile the new solar abundances with helioseismology, but the results are not encouraging: it does not seem possible to obtain a solar model in good agreement with the seismic inversions, contrary to models computed with the old abundances.

\bibliographystyle{aa}

\end{document}